\documentclass[aps,showpacs,twocolumn]{revtex4}
\usepackage{amsmath}
\usepackage{amssymb}
\usepackage{amscd}
\usepackage{wasysym}
\usepackage[ansinew]{inputenc}
\usepackage[T1]{fontenc}
\usepackage{ae,aecompl}

    \usepackage[dvips]{graphicx}
    \DeclareGraphicsExtensions{.eps}

\newcommand{\be}{\begin{equation}}
\newcommand{\ee}{\end{equation}}
\newcommand{\bea}{\begin{eqnarray}}
\newcommand{\eea}{\end{eqnarray}}
\newcommand{\nn}{\nonumber}

\begin{document}
\bibliographystyle{apsrev}

\title{On the effective Dirac dynamics of ultracold atoms in bichromatic optical lattices}
\author{D. Witthaut}
\affiliation{Max-Planck-Institute for Dynamics and Self-Organization,
D--37073 G\"ottingen, Germany}

\author{T. Salger}
\affiliation{Institut f\"ur Angewandte Physik,
Universit\"at Bonn,
D--53115 Bonn, Germany}

\author{S. Kling}
\affiliation{Institut f\"ur Angewandte Physik,
Universit\"at Bonn,
D--53115 Bonn, Germany}

\author{C. Grossert}
\affiliation{Institut f\"ur Angewandte Physik,
Universit\"at Bonn,
D--53115 Bonn, Germany}

\author{M. Weitz}
\affiliation{Institut f\"ur Angewandte Physik,
Universit\"at Bonn,
D--53115 Bonn, Germany}

\date{\today }

\begin{abstract}
We study the dynamics of ultracold atoms in
tailored bichromatic optical lattices.
By tuning the lattice parameters, one can readily engineer
the band structure and realize a Dirac point, 
i.e. a true crossing of two Bloch bands.
The dynamics in the vicinity of such a crossing is described by 
the one-dimensional Dirac equation, which is rigorously shown 
beyond the tight-binding approximation. 
Within this framework we analyze the effects of an
external potential and demonstrate numerically that it is possible 
to demonstrate Klein tunneling with current experimental setups.

\end{abstract}

\maketitle

\section{Introduction}

Quantum simulators aim at the simulation of complex quantum
systems in well controllable laboratory experiments \cite{Bulu09}.
Such a simulation is especially useful when the original quantum 
system is experimentally not accessible and numerical simulations 
are impossible due to the exponential size of the Hilbert space. 
Furthermore, quantum simulators offer the possibility to tune 
the experimental parameters to explore novel physical phenomena. 
Important examples include the simulation of solid state systems 
with ultracold atoms \cite{Bloc08}, sonic black holes in Bose-Einstein 
condensates \cite{Gara00} and the Dirac dynamics with trapped ions 
\cite{Lama07,Gerr10,Casa10,Gerr11}.

Ultracold atoms in optical lattices are especially suited for such 
a task, since their dynamics can be controlled with an astonishing
precision and their dynamics can be measured in situ.
Bichromatic lattices are especially appealing since these
systems allow to tune the energy dispersion of the Bloch
bands. In particular one can choose the parameters such
that a Dirac point, i.e. a true crossing of the first and second 
excited band, is realized depending on the relative phase
between the two fundamental lattices. 
Unlike other systems \cite{Juze08,Otte09,Witt10}, bichromatic
optical lattices thus allow to simulate relativistic quantum effects
using only a single species of neutral atoms and no external
driving fields. 
This approach therefore paves the way for the simulation of 
interacting relativistic quantum field theories \cite{Cira10}.

In this paper we investigate the dynamics around a Dirac 
point in detail and derive the one-dimensional Dirac equation 
as an effective equation of motion for the coarse-grained
atomic wave functions. In contrast to previous approaches
\cite{Zhu07,Apaj10,Long10,Cira10}, we do \emph{not} make use 
of a tight-binding approximation, such that the Dirac equation
is found without imposing a continuum limit.
We discuss the effects of an external potential in detail,
showing that the Dirac description remains valid if the
potential varies slowly enough.
Within this framework we finally show that it is possible to
simulate Klein tunneling through a potential barrier with 
current experimental methods.

\section{Bloch and Wannier states in bichromatic optical lattices}

We consider the dynamics of ultracold atoms in a bichromatic
optical lattice described by the Hamiltonian
\be
  \hat H_0 =  \frac{-\hbar^2}{2M} \frac{\partial^2 }{\partial x^2} 
       + \frac{V_1}{2} \cos(2k_0 x)   + \frac{V_2}{2} \cos(4k_0 x + \phi)
\ee
plus an additional potential $ V(x)$, which is assumed to vary
slowly compared to $k_0 x$. A bichromatic optical lattice with arbitrary 
relative phase $\phi$ can be implemented by a superposition
of an ordinary optical lattice with a periodicity of $\lambda/2$
and an additional lattice with a periodicity of $\lambda/4$ based on four-photon
processes as described in \cite{Ritt06,Salg07,Salg09}. For the additional 
potential $V(x)$ we consider (i) an optical dipole trap and (ii) a static 
field which can be realized by either gravity or accelerating
the complete lattice. 
In the following we will use scaled units which are obtained by setting 
$x' = k_0 x$, $t' = E_R t/\hbar$ and dividing the Schr\"odinger 
equation by the recoil energy $E_R = \hbar^2 k_0^2/2M$.
In these units we have $\hbar = 1$ and $M = 1/2$ and all 
energies are given in units of $E_R$. 

\begin{figure}[tb]
\centering
\includegraphics[width=8cm, angle=0]{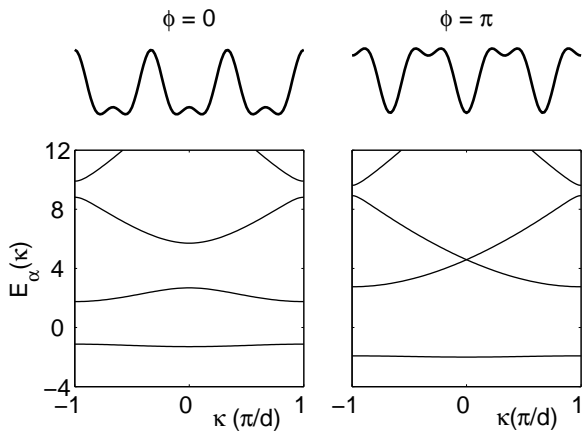}
\caption{\label{fig:diracbands}
The lowest Bloch bands for the case $V_1 = 5$ and $V_2 = 1.56$.
For $\phi = \pi$ one observes a true crossing between the first 
and the second excited band, a Dirac point, which can be 
used to simulate the Dirac equation.
}
\end{figure}

The eigenstates of $\hat H_0$, the Bloch waves can be engineered 
to a large extend by choosing the lattice parameters $V_{1,2}$ and 
$\phi$ which makes bichromatic lattices a convenient tool
for quantum simulations. To be precise, the Bloch states are 
defined as the simultaneous eigenstates of $\hat H_0$
and the translation $T_d$ over the lattice period $d$:
\bea
  && \hat H_0 u_{\alpha,\kappa}(x) = E_\alpha(\kappa)
  u_{\alpha,\kappa}(x), \nn \\
  && T_d u_{\alpha,\kappa}(x) = e^{i d \kappa}
  u_{\alpha,\kappa}(x).
  \label{eqn:bloch-def}
\eea
Here and in the following, $\kappa \in [-\pi/d,\pi/d]$ is the quasi 
momentum and $\alpha = 0,1,2,\ldots$ labels the different Bloch bands. 
Figure \ref{fig:diracbands} shows the bandstructure $E_{\alpha}(\kappa)$ 
of a bichromatic lattice with $V_1 = 5$ and $V_2 = 1.56$, comparing two 
different choices of the relative phase $\phi$.
For these values of the lattice depth and a relative phase
$\phi = \pi$, one observes a true crossing of the eigenenergies 
of the first and second exited band at $\kappa = 0$. i.e. a so-called
'Dirac point'. 
The physical reason for the vanishing of the band gap
is that the contributions of the second order Bragg scattering at the 
optical lattice with periodicity $\lambda/2$ and the
first order  Bragg scattering at the lattice with periodicity 
$\lambda/4$ show a complete destructive interference 
for the given parameters. 
At the Dirac point the dispersion relation is linear in $\kappa$,
just as for a relativistic massless Dirac particle, such that a similar
dynamics can be expected. We will make this analogy more precise 
in the following. In general, the band gap between the first and second
excited band is approximately given by 
$\Delta E \approx | (V_1/4)^2 + V_2 \exp(i \phi) |$ \cite{Salg07}.

\begin{figure}[tb]
\centering
\includegraphics[width=8cm, angle=0]{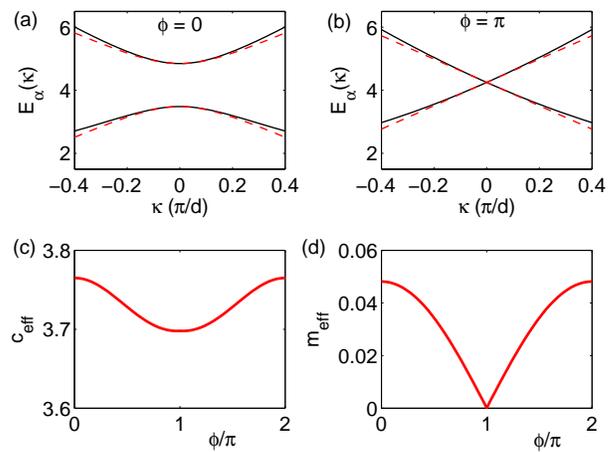}
\caption{\label{fig:diracfit}
(Color online)
(a,b) A fit of the relativistic dispersion relation (\ref{eqn:diracfit})
to the Bloch bands $\alpha=1,2$ in the center of the Brillouin zone
for $\phi = 0$ and $\phi = \pi$, respectively.
(c,d) The resulting fit values for the effective parameters
$m$ and $c$ as a function of the phase $\phi$.
The remaining parameters are the same 
as in Fig.~\ref{fig:diracbands}.
}
\end{figure}

However, also for a small but finite band gap we obtain a 
pseudo-relativistic dynamics in the center of the Brillouin 
zone. In any case, we can approximate the energy dispersion
of the first and second excited band around $\kappa \approx 0$
as 
\be
  E_{1,2}(\kappa) = E_D \pm \sqrt{m^2 c^4 + c^2 \kappa^2}.
  \label{eqn:diracfit}
\ee
This relation defines an effective mass $m$ which is given by
the curvature of the two bands and an effective speed of light 
$c$ which is related to the band gap by
\be
  \Delta E = 2 m c^2.
\ee
The applicability of this approximation is illustrated in 
Fig.~\ref{fig:diracfit}, where it is compared to the numerically 
exact data. Furthermore, the figure shows the effective 
parameters $m$ and $c$ as a function of the lattice phase $\phi$.

Now it is very convenient to introduce a new basis in which 
the two Bloch states in the first and second excited band are
rotated:
\bea
  \tilde u_{1,\kappa} &=& \cos \theta \, u_{1,\kappa}  +  \sin \theta \, u_{2,\kappa} 
   \nn \\
  \tilde u_{2,\kappa} &=& - \sin \theta \, u_{1,\kappa}  +  \cos \theta \,  u_{2,\kappa} 
  \label{eqn:basischange}
\eea
with the mixing angle
\be
  \tan \theta(\kappa) = \frac{mc^2}{c\kappa + \sqrt{m^2c^4 + c^2 \kappa^2}} .
\ee
In this basis, the free Hamiltonian $\hat H_0$ is no longer diagonal,
but has the convenient form
\be
 \hat H_0(\kappa) = 
  \begin{pmatrix} E_D + c \kappa & mc^2 \\
         mc^2 & E_D - c \kappa \end{pmatrix}
\ee
The eigenenergies (\ref{eqn:diracfit}) for given quasimomentum
$\kappa$ are simply the eigenvalues of the matrix $\hat H_0(\kappa)$.

\begin{figure}[tb]
\centering
\includegraphics[width=8cm, angle=0]{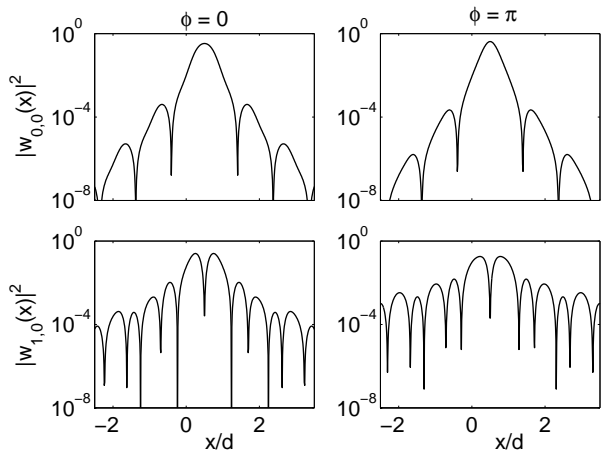}
\caption{\label{fig:wannier1}
Squared modulus of the Wannier states $|w_{\alpha,n}(x)|^2$ in
the ground ($\alpha = 0$, upper panels) and the first excited 
band ($\alpha = 1$, lower panels) at the lattice site $n=0$
in a semi-logarithmic scale.
We assume a bichromatic optical lattice with $V_1 = 5$, 
$V_2 = 1.56$ and $\phi = 0$ (left) and $\phi = \pi$, respectively.
}
\end{figure}

For an effective description of the quantum dynamics in a perturbed
crystal we will furthermore need the Wannier basis which is defined 
as follows. The Bloch waves can be chosen to be periodic in the 
quasi momentum $\kappa$, such that they can be expanded into a 
Fourier series,
\be
  u_{\alpha,\kappa}(x) = \frac{d}{\sqrt{2\pi}} \sum_{n \in \mathbb{Z}}
  e^{i \kappa d n} w_{\alpha,n}(x),
  \label{eqn:wannier-def}
\ee
which defines the Wannier states $w_{\alpha,n}(x)$. Inverting
the Fourier series yields
\be
  w_{\alpha,n}(x) = \frac{1}{\sqrt{2\pi} d} \int e^{-i n d  \kappa} 
      u_{\alpha,\kappa}(x) \, d\kappa.
     \label{eqn:wannier-int} 
\ee
The Wannier
states are exponentially localized at the lattice site $n$ \cite{Kohn59}, 
and they are related by a simple shift in real space
\be
  w_{\alpha,n}(x) = w_{\alpha,0}(x - x_n),
  \label{eqn:wannier-shift}
\ee
where $x_n = nd$ is the position of the $n$th lattice well.
These properties are quite useful for several approximation 
schemes  (cf. \cite{Jaks98}). 
Examples for the case of a bichromatic lattice 
are shown in Fig.~\ref{fig:wannier1} for two values of the
relative phase $\phi$.

\section{Effective evolution equations}

For the derivation of the effective evolution equations we start
from the Wannier representation of an arbitrary wave function,
\be
  \Psi(x) = \sum_{\alpha,n}  \psi_{\alpha,n}  w_{\alpha,0}(x - x_n)
\ee
The expansion coefficients $\psi_{\alpha,n}$ form an infinite,
but countable set of complex number. For the following 
approximations, however, we define a continuous function
$\psi_{\alpha}(x)$ for every Bloch band $\alpha$ such that
\be
  \psi_{\alpha}(x_n) = \psi_{\alpha,n}.
\ee
These functions can be viewed as a coarse grained version
of the quantum state $\Psi(x)$ projected on the band $\alpha$.
Now one can derive effective evolution equations for these
coarse grained wave functions.

One can show that every operator, which is diagonal in
the quasi momentum $\kappa$, can be expressed in a very 
convenient way for the coarse grained wave functions 
$\psi_{\alpha}(x)$.
So consider an operator which satisfies
\be
  \hat O \, u_{\alpha,\kappa} = \sum_{\beta} 
      O_{\beta,\alpha}(\kappa) u_{\beta,\kappa}
\ee
where $u_{\alpha,\kappa}$ are the Bloch states defined above.
Then this operator acts on the coarse grained wave function as
\be
  \hat O \psi_{\alpha}(x) = \sum_{\beta} 
  O_{\beta,\alpha}( \hat p )  \, \psi_{\beta}(x),
\ee
i.e. the quasi momentum $\kappa$ is replaced by the momentum
operator $\hat p = -i  \partial_x$. To carry out this replacement one can expand
$O_{\beta,\alpha}(\kappa)$ in a Taylor series and then replace
every term $\kappa^n$ by $(-i \partial_x)^n$. This relation was 
first shown by Slater \cite{Slat49}, cf. also \cite{Lutt51,Adam52,Sund99}.
The proof is summarized in the appendix.

In particular, this holds for the lattice Hamiltonian $\hat H_0$,
which is trivially diagonal in the Bloch basis.
Going to the rotated basis defined in equation 
(\ref{eqn:basischange}),
the eigenvalue equation for the lattice Hamiltonian reads
\be
  \hat H_0 
     \begin{pmatrix} \tilde u_{1,\kappa} \\ \tilde u_{2,\kappa} \end{pmatrix}
    = \begin{pmatrix} E_D + c \kappa & mc^2 \\
         mc^2 & E_D - c \kappa \end{pmatrix}
      \begin{pmatrix} \tilde u_{1,\kappa} \\ \tilde u_{2,\kappa} \end{pmatrix}        
\ee
The constant energy offset $E_D$ introduces a global phase shift only,
which has no physical significance. By shifting the energy scale, we
can set it to zero, $E_D = 0$.

Now if we express a general quantum state in the first and second
excited band as
\be
  \Psi(x) = \sum_{\alpha=1,2;n}  \psi_{\alpha}(x_n)  \tilde w_{\alpha,n}(x),
  \label{eqn:ansatz-wtilde}
\ee
where the rotated Wannier function $\tilde w_{\alpha,n}(x)$ are 
related to the original ones $w_{\alpha,n}(x)$  by the same
rotation as the used in Eqn.~(\ref{eqn:basischange}),
then Slater's theorem tells us that the lattice Hamiltonian acts
onto the coarse grained wave functions $\psi_{\alpha}(x)$ as
\be
  \hat H_0 
     \begin{pmatrix} \psi_1(x) \\ \psi_2(x) \end{pmatrix}
    = \begin{pmatrix} + c \hat p & mc^2 \\
         mc^2 & - c \hat p \end{pmatrix}
      \begin{pmatrix} \psi_1(x) \\ \psi_2(x) \end{pmatrix},      
   \label{eqn:h0-psi}  
\ee
where $\hat p = -i \partial_x$ is the momentum operator.

\begin{figure}[tb]
\centering
\includegraphics[width=8cm, angle=0]{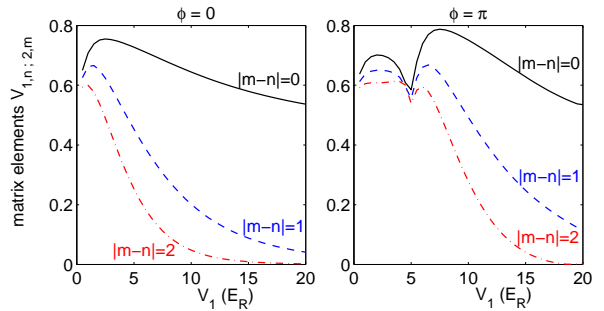}
\caption{\label{fig:matrixel}
(Color online)
Matrix elements $V_{\alpha,n; \beta,m}$ of a linear potential
$V(x) = x$ in a Wannier basis as a function of the lattice depth 
$V_1$. We have plotted  the matrix elements for $\alpha = 1$, 
$\beta  = 2$ and $|n-m| \leq 2$, which are the leading corrections 
to the diagonal approximation (\ref{eqn:pot_firststorder}).
As above, we assume $V_2/V_1 = 1.56/5$ and $\phi = 0$ (left) 
and $\phi = \pi$ (right), resepectively.
}
\end{figure}

In addition we need to know how a slowly varying potential
$V(x)$ acts onto the functions $\psi_{\alpha}(x)$. We recall that
these functions have been defined as expansion coefficients 
in the Wannier basis, such that we need the matrix elements
\be
  V_{\beta,n;\alpha,m} = \int dx \, 
  w^*_{\beta,0}(x - x_m) V(x)  w_{\alpha,0}(x - x_n).
\ee
Using the strong localization of the Wannier states, one can set
$V(x)$ to a constant over the localization length in a first 
approximation. Because of the orthogonality of the Wannier
states one thus finds  
\be
  V_{\beta,n;\alpha,m} \approx V(x_n) 
     \delta_{\beta,\alpha} \delta_{n,m}.
     \label{eqn:pot_firststorder}
\ee

\begin{figure*}[tb]
\centering
\includegraphics[height=7cm, angle=0]{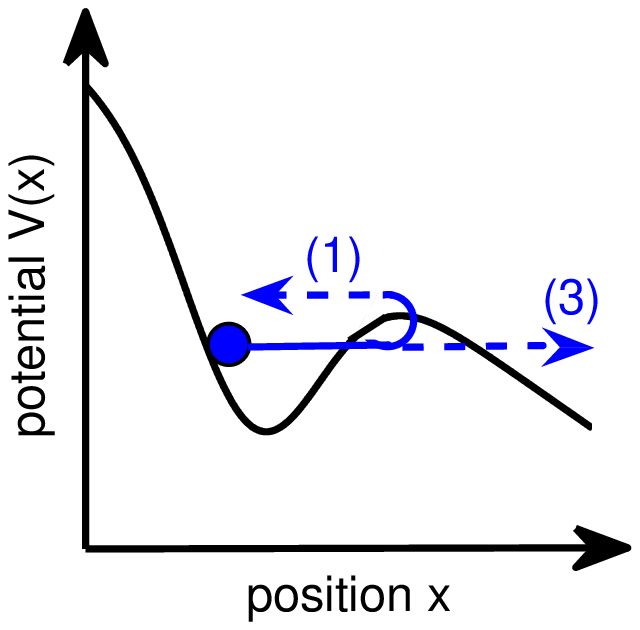}
\includegraphics[height=7cm, angle=0]{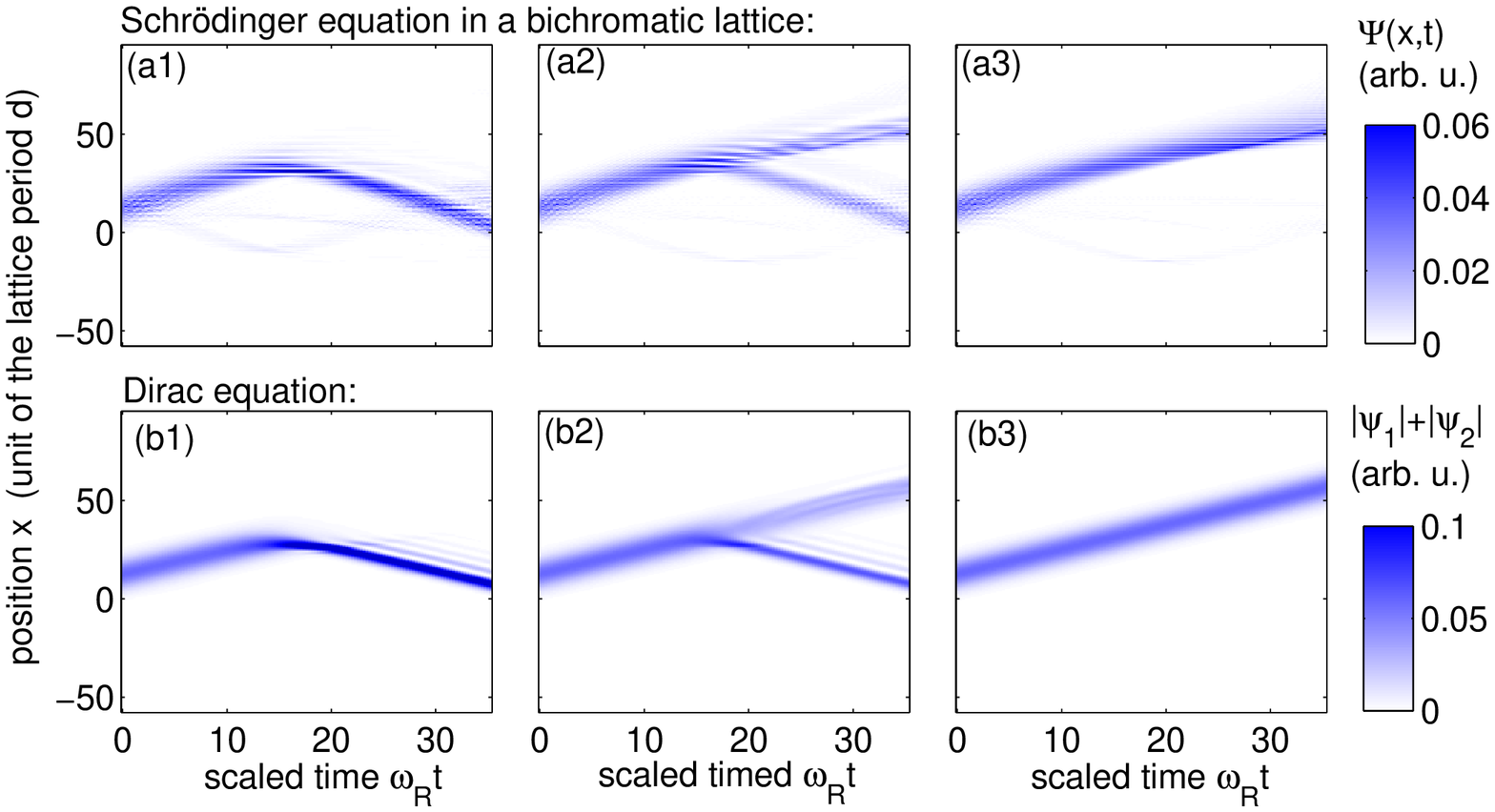}
\caption{\label{fig:klein1}
(Color online)
Quantum simulation of Klein-Tunneling out of a dipole trap.
The upper panels (a1-a3) show the full time evolution of the 
Schr\"odinger equation in a bichromatic optical lattice for the 
case $V_1 = 5$ and $V_2 = 1.56$ and a relative phase of 
$\phi = 0$ (a1), $\phi = 0.8 \, \pi$ (a2), and $\phi = \pi$ (a3).
The lower panels (b1-b3) show the dynamics of the effective 
Dirac equation (\ref{eqn:dirac1}) with effective mass and 
speed of light given by $mc^2 = 0.78$ (b1), $mc^2 = 0.24$ (b2), 
and $mc^2 = 0$ (b3), respectively.
}
\end{figure*}

Let us analyze this approximation in more detail for the case
of a linear potential $V(x) = Fx$. Then we have
\be
  V_{\beta,n;\alpha,m} = F x_n  \delta_{\beta,\alpha} \delta_{n,m}
   + F  \int dx \,  w^*_{\beta,m-n}(x) \, x \, w_{\alpha,0}(x), \nn
\ee
where Eqn.~(\ref{eqn:wannier-shift}) has been used to simplify
the results. The first term in this expression corresponds to the
diagonal approximation (\ref{eqn:pot_firststorder}), while the
remaining matrix elements can be interpreted as follows.
The local terms, i.e. the terms with $n=m$, vanish exactly
for $\alpha = \beta$ due to the parity of the Wannier functions.
The term $\alpha = 1$ and $\beta = 2$ is the most important
correction to the diagonal approximation (\ref{eqn:pot_firststorder}).
It couples the two bands and can thus be viewed as an additional
contribution to the effective mass $mc^2$ in the effective wave
equation (\ref{eqn:h0-psi}).  
The nonlocal terms $n \neq m$ are also largest for 
$\alpha = 1$ and $\beta = 2$. However, they vanish 
exponentially with the lattice depth $V_{1,2}$ and 
the squared lattice period $d^2$. 
These matrix elements are plotted as a function of the
lattice depth in Fig.~\ref{fig:matrixel}, assuming 
$V_2/V_1 = 1.56/5$ and $\phi = 0,\pi$ as above.
Note that the lattice depth in scaled units is proportional 
to $d^2$. One clearly sees how the non-local terms
vanish exponentially in contrast to the local term $n=m$.

In the following we will confine ourselves to the first order
approximation (\ref{eqn:pot_firststorder}). For the ansatz 
(\ref{eqn:ansatz-wtilde}), the potential thus acts as
\be
    V(x) \sum_{\alpha,n}  \psi_{\alpha}(x_n)  \tilde w_{\alpha,n}(x) 
    \approx \sum_{\alpha,n}  V(x_n)  \psi_{\alpha}(x_n)  \tilde w_{\alpha,n}(x).
    \nn
\ee
In this approximation, we thus find the effective evolution equations 
for the coarse grained wave function:
\be
  i \frac{\partial}{\partial t}
     \begin{pmatrix} \psi_1 \\ \psi_2 \end{pmatrix}
    = \begin{pmatrix} V(x) + c \hat p & mc^2 \\
         mc^2 & V(x) - c \hat p \end{pmatrix}
      \begin{pmatrix} \psi_1 \\ \psi_2 \end{pmatrix}.      
     \label{eqn:dirac1} 
\ee
If we rotate the 'spinor' wave function $(\psi_1,\psi_2)$ 
once again by the unitary transformation
\be
  U = \frac{1}{\sqrt{2}} \begin{pmatrix} 1 & -1 \\ 1 & 1 \end{pmatrix},
\ee
we finally obtain the Dirac equation in 1+1 dimensions with
an external scalar potential,
\be
  i \frac{\partial}{\partial t}
     \begin{pmatrix} \psi_a \\ \psi_b \end{pmatrix}
    = \begin{pmatrix} V(x) -mc^2 & c \hat p \\
         c \hat p & V(x) + mc^2 \end{pmatrix}
      \begin{pmatrix} \psi_a \\ \psi_b \end{pmatrix}.      
\ee
Note that in this rotated frame $\psi_a$ and $\psi_b$
coincide with the amplitudes in the first and second excited
band for $\kappa = 0$ -- but only there. Otherwise one
has to be very careful when interpreting the wave functions.

\section{Quantum simulation of Klein tunneling}

As an example of the effective relativistic dynamics we consider 
the tunneling of a wavepacket out of a dipole trap in a tilted 
bichromatic lattice. In particular, the slowly varying potential is 
given by
\be
  V(x) = -V_0 \exp(-2x^2 / W_0^2) -Fx
\ee
with $F = 0.076$, $V_0 = 19.77$ and $W_0 = 157$ in scaled
units, which corresponds to typical experimental parameters 
\cite{Salg07,Salg09}.
Initially, the wave packet is localized in the second
exited band with quasi momentum $\kappa = 0.95$
with a gaussian envelope of width $\sigma = 17$.

The resulting dynamics of the atoms is shown in Fig.~\ref{fig:klein1}. 
The sketch on the left demonstrates the two different regimes 
realized for $\phi = 0$ (a1, b1) and $\phi = \pi$ (a3, b3). In the
latter case a Dirac point emerges in the band structure,
such that the atoms behave like massless relativistic particles.
Unlike a massive Schr\"odinger particle such a Dirac particle
can escape from the trap via \emph{Klein tunneling}.

This expectation is confirmed by the numerical simulation of
the atomic dynamics. The upper panels (a1-a3) of Fig.~\ref{fig:klein1}
show the evolution of the modulus of the atomic wave function 
$|\Psi(x,t)|$, calculated with the original Schr\"odinger equation
in a bichromatic optical lattice.
The effective Dirac dynamics according to Eqn.~(\ref{eqn:dirac1})
is shown in the lower panels (b1-b3). The effective values of the
mass $m$ and the speed of light $c$ are given by
$mc^2 = 0.78$ (b1), $mc^2 = 0.24$ (b2), and
$mc^2 = 0$ (b3), respectively.
For the sake of a better visibility we have again plotted 
the modulus $|\psi_1(x,t)| + |\psi_2(x,t)|$.
One observes that the essential features of the atomic dynamics 
are very well reproduced by the Dirac approximation. In particular,
one observes the transition from a 'heavy' Schr\"odinger-like
particle for $\phi = 0$ (a1, b1) to a 'relativistic' particle with 
vanishing effective mass for $\phi = \pi$ (a3, b3), which 
escapes from the dipole trap by Klein tunneling. 
For $\phi = 0.8 \, \pi$ (a2, b2), 
partial tunneling is observed and the potential barrier 
acts as a matter wave beam splitter.

The most obvious difference of the Dirac approximation 
(\ref{eqn:dirac1}) to the underlying lattice dynamics
is that the group velocity of the Dirac wave packet is 
limited to the effective speed of light $c$. After tunneling
out of the dipole trap, the atoms are accelerated by the 
linear potential $Fx$, which is not observed in an effective
relativistic description.
This difference is due to the fact that the wave packet is
not restricted to the center of the Brillouin zone for the
given parameters, as it was assumed in the derivation 
of the effective Dirac equation. Instead, we have
$\kappa_{\rm initial} = 0.9$ and $|\kappa_{\rm final}| \apprle 1$
in the example shown in Fig.~\ref{fig:klein1}.

\section{Conclusion and Outlook}

We have analyzed the quantum dynamics of ultracold
atoms in a bichromatic optical lattice. It was shown that
the lattice parameters can be tuned such that a Dirac
point emerges in the band structure, i.e. a true crossing 
of the Bloch bands with linear dispersion relation. 
In the vicinity of such a crossing the atoms effectively 
behave like massless relativistic particles, allowing
for a tabletop simulation of relativistic quantum physics.

We have rigorously shown that the one-dimensional Dirac 
equation is found as an effective evolution equation for
the coarse grained atomic wave function projected onto
the two crossing Bloch bands. Unlike previous approaches,
our derivation does not rely on a tight-binding approximation
and also shows that how to include an additional, slowly 
varying potential. Therefore it is possible to simulate 
\emph{Klein tunneling} -- the tunneling of an ultrarelativistic
particle though a potential barrier without damping --
with current experimental setups. 

Ultracold atoms in bichromatic optical lattices have some
important advantages than other systems proposed before.
The initial state of the atoms and all experimental parameters
can be controlled with astonishing precision. 
In a common experiment, several thousands of ultracold atoms 
are prepared in the optical lattice. Thus it is possible to simulate
the dynamics of \emph{interacting} Dirac fermions in contrast
to experiments with single trapped ions.
While these experiments of course require the use of 
fermionic ultracold atoms, we note that Klein tunneling
is a single particle effect such that it can be observed 
equally well for bosonic  atoms.

\begin{acknowledgments}

We thank A.~Rosch, H.~Kroha and K.~Ziegler for stimulating
discussions. Financial support by the German Research 
Foundation (DFG) and the Max Planck society is gratefully
acknowledged.

\end{acknowledgments}

\appendix*

\section{Slater's derivation of the effective equation of motion}

The wave function $\Psi(x,t)$ is expanded into Wannier states
\bea
  \Psi(x,t) &=& \sum_{\alpha,n} \psi_{\alpha,x} w_{\alpha,n}(x) \nn \\
      &=& \sum_{\alpha,n} \psi_{\alpha}(x_n) w_{\alpha,0}(x-x_n). 
     \label{eqn:ws-expand} 
\eea
The wave function is thus represented by a discrete set of numbers 
$\psi_{\alpha,n}$. Due to the exponential localization of the Wannier
states, these coefficients can be interpreted as the amplitude in the 
$n$th lattice site and in band $\alpha$.
For the effective evolution equations treat these coefficients as a
continuous function in $x$, i.e. we choose a smooth function 
$\psi_{\alpha}(x)$ such that
\be
  \psi_{\alpha}(x_n) = \psi_{\alpha,n}.
\ee
This function can be seen as a coarse grained version of the
original wave function $\Psi(x)$ projected onto the $\alpha$th
Bloch band.

We consider an operator, which is diagonal in the quasi 
momentum $\kappa$,
\be
  \hat O \, u_{\alpha,\kappa} = \sum_{\beta} 
      O_{\beta,\alpha}(\kappa) u_{\beta,\kappa},
\ee
as for instance the unperturbed Hamiltonian $\hat H_0$.
The functions $O_{\beta,\alpha}(\kappa)$ are periodic
in $\kappa$, such that we can expand them
into a Fourier series:
\be
  O_{\beta,\alpha}(\kappa) = \sum_s O_{\beta,\alpha}^{(s)} e^{-is d\kappa}.
\ee
Applying this operator to a wave function of the form 
(\ref{eqn:ws-expand}) and inserting the definition
(\ref{eqn:wannier-int}) for the Wannier function yields 
yields
\bea
  && \!\! \hat O \Psi(x) = \sum_{\alpha,n}  \psi_{\alpha}(x_n)  \,
         \hat O \, w_{\alpha,n}(x) \nn \\
  && \; = \frac{1}{\sqrt{2\pi} d}  \int d\kappa 
   \sum_{\alpha,\beta,n,s} \psi_{\alpha}(x_n) 
     O_{\beta,\alpha}^{(s)}  e^{-i (n+s) d\kappa} 
       u_{\beta,\kappa}(x). \nn
\eea
Using again equation (\ref{eqn:wannier-int}) this can be rewritten
as
\bea
   \hat O \Psi(x) &=&   \sum_{\alpha,\beta,n,s} \psi_{\alpha}(x_n) 
        O_{\beta,\alpha}^{(s)} w_{\beta}(x - x_{n+s}) \nn \\
       &=& \sum_{\alpha,\beta,m,s} 
     O_{\beta,\alpha}^{(s)} \psi_\alpha(x_m - x_s) w_{\beta}(x - x_m), \nn
\eea
where we have set $x_m = x_n + x_s$.
Now one can use the the spatial translation operator and fact that 
we assumed $\psi_\alpha(x)$ to be a continuous function 
such that
\be
   \psi_\alpha(x_m - x_s) = e^{-i sd \hat p} \psi_\alpha(x_m),
\ee
where $\hat p = -i \partial_x$ is the momentum operator.
We then finally obtain
\bea
  && \hat O  \sum_{\alpha,m} 
       \psi_\alpha(x_m) w_{\alpha,m}(x)  \nn \\
  && \qquad =  \sum_{\alpha,\beta,m}  \sum_{s} 
     O_{\alpha,\beta}^{(s)} e^{-i sd \hat p} \;
         \psi_\beta(x_m) w_{\alpha}(x - x_m) \nn \\
  && \qquad = \sum_{\alpha,\beta,m} 
     O_{\alpha,\beta}(\hat p) \psi_\beta(x_m) w_{\alpha,m}(x).
\eea
Comparing coefficients we find the desired relation
\be
  \hat O  \psi_\alpha(x)
    =  \sum_\beta O_{\alpha,\beta}(\hat p) \; \psi_\beta(x).
\ee



\begin{thebibliography}{10}

\bibitem{Bulu09}
I.~Bulutaand and F.~Nori,
Science {\bf 326}, 106 (2009).

\bibitem{Bloc08}
I.~Bloch, J.~Dalibard, and W.~Zwerger,
Rev. Mod. Phys. {\bf 80}, 885 (2008). 

\bibitem{Gara00}
L.~J. Garay, J.~R. Anglin, J.~I. Cirac, and P.~Zoller,
Phys. Rev. Lett. {\bf 85}, 4643 (2000).

\bibitem{Lama07}
L.~Lamata,  J.~Le\'on, T. Sch\"atz, and E.~Solano, 
Phys. Rev. Lett. {\bf 98}, 253005 (2007).

\bibitem{Gerr10}
R.~Gerritsma, F. Z\"ahringer, E.~Solano, R.~Blatt, and C.~F. Roos,
Nature {\bf 463},  68 (2010).

\bibitem{Casa10}
J.~Casanova, J.~J. Garcia-Ripoll, R.~Gerritsma, C.~F. Roos, and E.~Solano,
Phys. Rev. A \textbf{82}, 020101 (2010).

\bibitem{Gerr11}
R.~Gerritsma, B.~Lanyon, G.~Kirchmair, F.~Z\"ahringer, C.~Hempel, 
J.~Casanova, J.~J. Garcia-Ripoll, E.~Solano, R.~Blatt, and C.~F. Roos, 
Phys. Rev. Lett. \textbf{106}, 060503 (2011).



\bibitem{Juze08}
G.~Juzeliunas, J.~Ruseckas, M.~Lindberg, L.~Santos, and P.~\"Ohberg,
Phys. Rev. A \textbf{77}, 011802(R) (2008).


\bibitem{Otte09}
J.~Otterbach, R.~G. Unanyan, and M. Fleischhauer,
Phys. Rev. Lett. {\bf 102}, 063602 (2009).

\bibitem{Witt10}
D. Witthaut,
Phys. Rev. A \textbf{82}, 033602 (2010).

\bibitem{Cira10}
J.~I. Cirac, P.~Maraner, and J.~K. Pachos,
Phys. Rev. Lett. \textbf{105}, 190403 (2010).

\bibitem{Zhu07}
S.-L.~Zhu, B.~Wang, and L.-M.~Duan,  
Phys. Rev. Lett. {\bf 98}, 260402 (2007).

\bibitem{Long10}
S. Longhi,
Phys. Rev. B \textbf{81}, 075102 (2010).

\bibitem{Apaj10}
V. Apaja, M. Hyrk\"as, and M. Manninen,
Phys. Rev. A \textbf{82}, 041402 (2010).

\bibitem{Ritt06}
G.~Ritt, C.~Geckeler, T.~Salger, G.~Cennini, and M.~Weitz, 
Phys.~Rev.~A {\bf 74}, 063622 (2006).

\bibitem{Salg07}
T.~Salger, C.~Geckeler, S.~Kling, and M.~Weitz,
Phys. Rev. Lett. {\bf 99}, 190405 (2007).

\bibitem{Salg09}
T.~Salger, C.~Geckeler, S.~Kling, and M.~Weitz,
Science {\bf 326}, 1241 (2009).

\bibitem{Bloc28}
F.~Bloch, Z. Phys. {\bf 52}, 555 (1928). 




\bibitem{Kohn59}
W.~Kohn,   Phys. Rev.  {\bf 115},     809 (1959).

\bibitem{Jaks98}
D.~Jaksch, C.~Bruder, J.~I. Cirac, C.~W. Gardiner, and P.~Zoller,
Phys. Rev. Lett. {\bf 81}, 3108  (1998).



\bibitem{Thal92}
B.~Thaller, {\it The Dirac eqation},
Springer, Berlin Heidelberg New York (1992). 



\bibitem{Slat49}
J.~C. Slater, Phys. Rev. {\bf 76}, 1592 (1949).

\bibitem{Lutt51}
J.~M. Luttinger, Phys. Rev. {\bf 84}, 814 (1951).

\bibitem{Adam52}
E.~N. Adams II, Phys. Rev. {\bf 85}, 41 (1952).


\bibitem{Sund99}
G.~Sundaram and Q.~Niu,
Phys. Rev. B {\bf 59}, 14915 (1999).







\end{thebibliography}
\end{document}